\documentclass{sig-alternate}

\usepackage[resetlabels,labeled]{multibib}
\newcites{S}{Primary Sources Cites}

\usepackage{floatrow}
\usepackage{amsfonts}
\usepackage{mathtools}
\usepackage{algorithmic}
\usepackage{algorithm}
\usepackage{graphicx}
\usepackage[tight,footnotesize]{subfigure}
\usepackage{cite}
\usepackage{multirow}

\usepackage{array}
\usepackage{multicol}
\usepackage{url}

\newfloatcommand{capbtabbox}{table}[][\FBwidth]

\usepackage{amssymb}
\usepackage{graphicx}

\usepackage{amsthm}
\usepackage{xspace}

\usepackage{soul}

\usepackage[table]{xcolor}
\xdefinecolor{gray95}{gray}{0.65}
\xdefinecolor{gray25}{gray}{0.8}

\newcommand{\numarticles}{42\xspace}

\newtheoremstyle{exampstyle}
{1pt} 
{1pt} 
{} 
{} 
{\bfseries} 
{.} 
{.5em} 
{} 

\theoremstyle{exampstyle}

\begin{document}
\clubpenalty=10000
\widowpenalty = 10000
    
%

\title{A Hitchhiker's  Guide to Search-Based Software Engineering for Software Product Lines}

\numberofauthors{6} 
%
%
%


\author{
\alignauthor Roberto E. Lopez-Herrejon\\
       \affaddr{Johannes Kepler University Linz, Austria}\\
       \email{rlopez@jku.at}
\alignauthor Javier Ferrer\\
       \affaddr{Universidad de M\'alaga, Andaluc\'{\i}a Tech, Spain}\\
       \email{ferrer@lcc.uma.es}
\alignauthor Francisco Chicano\\
       \affaddr{Universidad de M\'alaga, Andaluc\'{\i}a Tech, Spain}\\
       \email{chicano@lcc.uma.es}
\and  
\alignauthor Lukas Linsbauer\\
       \affaddr{Johannes Kepler University Linz, Austria}\\
       \email{lukas.linsbauer@jku.at}
\alignauthor Alexander Egyed\\
       \affaddr{Johannes Kepler University Linz, Austria}\\
       \email{alexander.egyed@jku.at}
\alignauthor Enrique Alba\\
       \affaddr{Universidad de M\'alaga, Andaluc\'{\i}a Tech, Spain}\\
       \email{eat@lcc.uma.es}
}

\maketitle

\begin{abstract}
Search Based Software Engineering (SBSE) is an emerging discipline that focuses on the application of search-based optimization techniques to software engineering problems. The capacity of SBSE techniques to tackle problems involving large search spaces make their application attractive for Software Product Lines (SPLs). 
In recent years, several publications have appeared that apply SBSE techniques to  SPL problems.
In this paper, we present the results of a systematic mapping study of such publications. We identified the stages of the SPL life cycle where SBSE techniques have been used, what case studies have been employed and how they have been analysed.
This mapping study revealed potential venues for further research as well as common misunderstanding and pitfalls when applying SBSE techniques that we address by providing a guideline for researchers and practitioners interested in exploiting these techniques.
\end{abstract}






\section{Introduction}
\label{sec:introduction}
\textit{Search Based Software Engineering (SBSE)} is an emerging discipline that focuses on the application of search-based optimization techniques to software engineering problems~\cite{DBLP:journals/csur/HarmanMZ12}. 
Among the techniques SBSE relies on are: evolutionary computation techniques\footnote{Evolutionary computation is an area of computer science, artificial intelligence more concretely, that studies algorithms that follow Darwinian principles of evolution~\cite{EibenSmith2003EC}.}(e.g. genetic algorithms) and basic local searches (e.g. hill climbing, simulated annealing or random search)~\cite{Luke2009Metaheuristics}. These techniques are generic, robust, and have been shown to scale to large search spaces. These capacities make their application attractive for \textit{Software Product Line (SPL)} problems. 

In recent years, several publications have appeared that explored applications of SBSE techniques to concrete SPL problems. This fact prompted us to carry out a systematic mapping study to provide an overview of this research area~\cite{conf/ease/Petersen08}. Our general goal is to identify the quantity and the type of research and results available, and thus highlight possible open research problems and opportunities. 
The research questions our study addresses are:
\begin{itemize}

\item \textbf{\textit{RQ1. In what phases of the SPL life cycle have SBSE techniques been used?}.} SBSE has been applied throughout the entire life cycle of single systems, so our interest is finding out if SBSE has or can be applied also throughout the entire life cycle of SPLs.

\item \textbf{\textit{RQ2. What SBSE techniques have been used?}.} There are a vast number of SBSE techniques available in literature. Our goal here is cataloguing their use for SPLs problems and analyse if there are common trends in their application.

\item \textbf{\textit{RQ3. What type of comparative analysis is used?}.} Search-based techniques commonly rely on randomness, so adequate statistical analysis is necessary for the results to be useful and meaningful. Here our objective is to gauge at the adequacy of this type of analysis depending on the techniques and problems used.

\item \textbf{\textit{RQ4. What evaluation case studies are used?}.} Here our focus is on cataloguing the type, number, and provenance of the case studies analysed. We believe that identifying common case studies and their sources could lead to establishing community-wide benchmarks for certain problems.  

\end{itemize}

Our study corroborated the increasing interest in applying SBSE techniques in SPLs.
We found that the most common application is testing at the Domain Engineering level, and the most common technique being genetic algorithms with an increasing interest in multi-objective optimization. We identified a need to improve the empirical evaluation with a more adequate statistical analysis, and some common pitfalls when dealing with multi-objective optimization algorithms, and provide a short guideline to address them.


\section{Systematic Mapping Study}
\label{sec:mapping-study}
\textit{Evidence-Based Software Engineering (EBSE)} is an emerging software engineering area whose goal is \textit{"to provide the means by which current best evidence from research can be integrated with practical experience and human values in the decision making process regarding the development and maintenance of software"}~\cite{kitchenhametal04}. One of the approaches advocated by EBSE is systematic mapping studies whose goal is to provide an overview of the results available within an area by categorizing them along criteria such as type, forum, frequency, etc.~\cite{conf/ease/Petersen08}. In this section we describe how we carried out the standard systematic mapping process and summarized the results we obtained.

\subsection{Process}
\label{subsec:process}

We carried out the standard five steps of systematic mapping studies as described by Petersen et al.~\cite{conf/ease/Petersen08}.

~\linebreak
\noindent \textbf{\textit{Step 1. Definition of Research Questions.}} These are the questions we put forward in Section~\ref{sec:introduction}. In short, our driving goal was to find out in what parts of the SPL life cycle which SBSE techniques have been used and to gauge the adequacy of the analysis performed and the types and provenance of case studies employed.

~\linebreak
\noindent \textbf{\textit{Step 2. Conduct Search for Primary Studies.}} As a first step in our search we selected the search terms and categorized them in SPL and SBSE terms. Table~\ref{tab:search-terms} shows the list of terms used. We should point out the search terms we used for SBSE are an extended version of the ones employed by Harman et al. in a recent survey of SBSE~\cite{DBLP:journals/csur/HarmanMZ12}. 

\begin{sloppypar}
To perform our search we proceeded in two stages. At the first stage, we performed queries in two specialized repositories, the Search Based Software Engineering Repository\footnote{\url{http://crestweb.cs.ucl.ac.uk/resources/sbse_repository/repository.html}},  and the Bibliography on Genetic Programming\footnote{\url{http://liinwww.ira.uka.de/bibliography/Ai/genetic.programming.html}}.
At the second stage, we relied on general search engines: ScienceDirect, IEEExplore, ACM Digital Libray, SpringerLink, and Google Scholar.  
The queries we performed took all the combinations of one term from the SPL list and one or more terms of the SBSE terms depending on the querying functionality of the search engines or repository. 
For example, the following is a query used in the IEEExplore engine:
\end{sloppypar}

\begin{center}
\begin{tabular}{p{8cm}}
\texttt{("product line") AND ("search-based" OR  
"search based" OR "optimization" ~OR~ "multi-objective optimization" OR "multiobjective optimization"  OR  "genetic algorithm" OR  "GA" OR ~"genetic ~programming" OR  "GP" OR  "hill climbing" OR  ~"simulated annealing")}
\end{tabular}
\end{center}

In addition, because SBSE is considered to have started with the seminal paper by Harman and Jones in 2001~\cite{DBLP:journals/infsof/HarmanJ01}, we trimmed our search to include only publications on or after that year.
We performed the pertinent queries from March 12th to March 31st, 2014. The queries yielded a total of 1,326 hits that we sieved as described next.

\begin{table}[h!]
    \begin{tabular}{|p{8cm}|}
    \hline
    \textbf{SPL terms:} product line, software family, feature model, variability, variant, commonality, variability-intensive system, highly-configurable system
\\
\\
    \textbf{SBSE terms:} search based, optimization, genetic algorithm (GA), multiobjective optimization, multi-objective optimization,  genetic programming (GP), hill climbing, simulated annealing, local search, integer programming, ant colony optimization (ACO), particle swarm optimization (PSO), scattered search, artificial immune systems (AIS), evolutionary algorithm, evolutionary strategy, greedy, greedy search, memetic algorithm, evolutionary programming, grammatical evolution, variable neighborhood search, iterative local search (ILS), GRASP, tabu search, path relinking, harmony search, imperial competitive algorithm, bee colony, fire fly, constraint handling, mutation testing
\\
    \hline
    \end{tabular}
  \caption{Summary of SPL and SBSE Search Terms}
  \label{tab:search-terms}
\end{table}

~\linebreak
\noindent \textbf{\textit{Step 3. Screening of Papers for Inclusion and Exclusion.}} This step was straightforward. We looked for the search terms it the title, abstract and keywords and whenever necessary at the introduction or at other places of the paper. The sole criteria for inclusion in our mapping study was that a clear application of SBSE techniques to SPL was described. This resulted in a total of \numarticles articles whose details are summarized in Table~\ref{tab:primary-sources}, presented in the order they were found.
We should point out that many of the hits were either applications of SBSE techniques to single software systems (i.e. not in the realm of SPLs), or for product lines in the general manufacturing or marketing sense but not for software.

~\linebreak
\textbf{\textit{Step 4. Paper Classification.}}
For the classification of the stages of the SPL life cycle we used Pohl et al.'s SPL engineering framework (see~\cite{SPLE}), which defines four sub-processes for both \textit{Domain Engineering (DE)} and \textit{Application Engineering (AE)}. We regard each sub-process as a stage. In addition to the eight stages of this framework, we considered two more classifications categories: one to cover all maintenance and evolution issues of SPLs, and one to contain publications that have tooling support as one of their main contributions. 
In summary, our classification terms are as follows and we will refer to them henceforth by their shorthand names in parenthesis:
\begin{itemize}
\item	\ul{\textit{Domain Requirements Engineering (DRE)}} is the sub-process of DE where the common and variable requirements of the product line are defined, documented in reusable requirements artefacts, and continuously managed.

\item	\ul{\textit{Domain Design (DD)}} is the sub-process of DE where a reference architecture for the entire software product line is developed.

\item	\ul{\textit{Domain Realisation (DR)}} is the sub-process of DE where the set of reusable components and interfaces of the product line is developed.

\item	\ul{\textit{Domain Testing (DT)}} is the sub-process of DE where evidence of defects in domain artefacts is uncovered and where reusable test artefacts for application testing are created.

\item	\ul{\textit{Application Requirements Engineering (ARE)}} is the sub-process of AE dealing with the elicitation of stakeholder requirements, the creation of the application requirements specification, and the management of application requirements.

\item	\ul{\textit{Application Design (AD)}} is the sub-process of AE where the reference architecture is specialised into the application architecture.

\item	\ul{\textit{Application Realisation (AR)}} is the sub-process of AE where a single application is realised according to the application architecture by reusing domain realisation artefacts.

\item	\ul{\textit{Application Testing (AT)}} is the sub-process of AE where domain test artefacts are reused to uncover evidence of defects in the application.

\item \ul{\textit{Maintenance and Evolution (ME)}} refers to all the maintenance and evolution of all the artefacts developed across the entire life cycle of SPLs. Reverse engineering artefacts or bug fixing are examples of activities that fall in this category.

\item \ul{Tool support (TOOL)} used to classify the publications that have tooling support as one of their main contributions (e.g. tool papers or demos). 

\end{itemize}

\textbf{\textit{Step 5. Data Extraction and Mapping Studies.}}
To perform the data extraction of our mapping study, the authors formed three distinct groups: one with more expertise in SBSE, one with more expertise in SPLs, and one with mixed background. Each group independently filled out a spreadsheet with the following data: \textit{i)} SPL Stage (as defined in Step 4), \textit{ii)} artefacts employed, \textit{iii)} rationale	for the categorization if any, \textit{iv)} SBSE techniques, \textit{v)} analysis performed (e.g. statistical test), \textit{vi)} number of case studies evaluated, \textit{vii)} type of case studies (e.g. artefact used), \textit{viii)} provenance of the case studies, and \textit{ix)} a general field for any remarks.  

Once the data was independently gathered, it was consolidated during a joint revision session of the three groups where the data of each article was discussed until a consensus was reached. A summary of the results obtained are shown in Table~\ref{tab:primary-sources}.
It should be pointed out that it is customary in SBSE articles to compare techniques against local or random searches. For our categorization, we report only the main SBSE technique put forward by each paper, regardless of other techniques used for comparison purposes.


\begin{table*}
\center
\caption{Primary Sources Summary Table}
\begin{tabular}{|c|p{4cm}|c|c|c|c|p{2cm}|c|}
\hline
\textbf{ID} & \textbf{Authors} & \textbf{Forum} & \textbf{Year} & \textbf{Technique} & \textbf{Stage} & \textbf{Analysis} & \textbf{Studies} \\ 
\hline
\citeS{DBLP:conf/gecco/WangAG13} &  Wang, Shaukat, Gotlieb & GECCO &	2013 & GA & DT & Stat.Tests & 6,FM$^{1,5}$\\ \hline
\citeS{DBLP:conf/ssbse/ColanziV12} &   Colanzi, Vergilio  & SSBSE &	2012 & MOEA & DD & None & 1,CD$^{4}$ \\ \hline
\citeS{DBLP:conf/ssbse/Lopez-HerrejonGBSE12} &  Lopez-Herrejon, Galindo, Benavides, Segura, Egyed & SSBSE & 2012 & GA & ME & Basic & 59,FM$^{1}$\\ \hline

\citeS{DBLP:conf/icse/SayyadMA13} &  Sayyad, Menzies, Ammar & ICSE & 2013 & MOEA &  ARE  & Basic & 2,FM$^{1}$\\ \hline

\citeS{DBLP:conf/icse/Colanzi12} &  Colanzi & ICSE & 2012 & MOEA & DD & None & None \\ \hline

\citeS{DBLP:journals/jss/GuoWWLW11} &   Guo, White, Wang, Li, Wang  & JSS &	2011 & GA & ARE & Basic & 900,FM$^{2}$\\ \hline

\citeS{conferences/ssbse/LopezHerrejon11} &  Lopez-Herrejon, Egyed & SSBSE &	2011 & DS  & ME & Basic & 60,FM$^{1}$\\ \hline

\citeS{DBLP:journals/ijitdm/WuTKC11} &   Wu,  Tang,  Kwong,  Chan & JISTDM &	2011 & IP & DRE & Undefined & 1,AH$^{5}$ \\ \hline

\citeS{DBLP:journals/tse/CohenDS08} &  Cohen,  Dwyer, Shi  & TSE &	2008 & GRE & DT & Undefined &  4,AH$^{3}$ \\ \hline

\citeS{Ullah09} &  Ullah & TR &	2009 & GA & ME & Undefined & 1,AH$^{3}$ \\ \hline

\citeS{DBLP:journals/ese/GarvinCD11} &  Garvin, Cohen, Dwyer  & ESE &	2011 & SA & DT  & Stat.Tests & 35,AH$^{2,3}$ \\ \hline

\citeS{DBLP:conf/issta/CohenDS07} &  Cohen,  Dwyer, Shi  & ISSTA &	2007 & GRE, SA & DT & Basic & 2,AH$^{3}$ \\ \hline

\citeS{DBLP:journals/eswa/SeguraPHBC14} & Segura, Parejo, Hierons, Benavides, Ruiz-Cort{\'e}s & ESA
& 2014 & GA & TOOL & Stat.Tests & 5000,FM$^{2}$ \\ \hline

\citeS{DBLP:conf/splc/HenardPPKT13a} & Henard, Papadakis, Perrouin, Klein, Le Traon
 & SPLC &	2013 & GA & TOOL, DT & None & None \\ \hline

\citeS{DBLP:conf/vamos/HaslingerLE13} &  Haslinger, Lopez-Herrejon, Egyed & VaMoS &	2013 & GRE  & DT & Basic & 146,FM$^{1}$ \\ \hline

\citeS{DBLP:conf/vamos/Lopez-HerrejonE12} &  Lopez-Herrejon, Egyed & VaMoS &	2012 & DS  & ME & Basic & 45,FM$^{1}$ \\ \hline

\citeS{DBLP:conf/splc/XuCMR13} &  Xu, Cohen,  Motycka, Rothermel & SPLC & 2013 & GA  & DT & Stat.Tests  & 2,FM$^{4}$ \\ \hline

\citeS{DBLP:conf/splc/HenardPPKT13} & Henard, Papadakis, Perrouin, Klein, Le Traon
 & SPLC &	2013 & GA & DT & Stat.Tests & 8,FM$^{1}$\\ \hline

\citeS{DBLP:conf/icse/HenardPPKT04} & Henard, Papadakis, Perrouin, Klein, Le Traon
 & ICSE &	2013 & LS & ME & Basic & 1,FM$^{1}$ \\ \hline

\citeS{Tan:2013:IPC:2525401.2525415} & Tan,  Lin, Ye, Zhang & ACSC
&	2013 & SA, GA & ARE & Basic & 1,FM$^{4}$ \\ \hline

\citeS{DBLP:conf/icse/PascualPF13} & Pascual, Pinto, Fuentes & SEAMS
&	2013 & GA & ARE, AR & Basic & 1,AH$^{4}$\\ \hline

\citeS{DBLP:conf/splc/SerajzadehS11} & Serajzadeh, Shams  & SPLC &	2011 & PSO & DRE & None & 1,AH$^{4}$ \\ \hline

\citeS{DBLP:conf/splc/JohansenHF12} & Johansen, Haugen, Fleurey  & SPLC &	2012 & GRE & DT & Basic & 19,FM$^{1,4}$ \\ \hline

\citeS{DBLP:conf/models/JohansenHFES12} & Johansen, Haugen, Fleurey, Eldegard, Syversen  & MODELS &	2012 & GRE & DT & None & 2,FM$^{3,5}$\\ \hline

\citeS{DBLP:conf/splc/MurashkinARC13} & Murashkin, Antkiewicz, Rayside, Czarnecki
  & SPLC  &	2013 & AEMOO & TOOL, ARE & None & None \\ \hline

\citeS{DBLP:conf/icse/ColanziV13} & Colanzi, Vergilio  & ICSE  &	2013 & MOEA & DD  & None & None \\ \hline

\citeS{DBLP:conf/icsm/Lopez-HerrejonCFEA13} & Lopez-Herrejon, Chicano, Ferrer, Egyed, Alba  & ICSM  &	2013 & AEMOO & DT & Basic & 118,FM$^{1}$ \\ \hline

\citeS{SEDM/Wu10} & Wu, Tang, Wang & SEDM  & 2010 & ADHS & DD, ARE & Undefined & None \\ \hline

\citeS{DBLP:conf/cec/CruzNBRASM13} & Cruz,Neto, Britto, Rabelo, Ayala, Soares, Mota & CEC  & 2013 & MOEA & ARE & Undefined & 1,AH$^{4}$\\ \hline

\citeS{DBLP:conf/kbse/SayyadIMA13} & Sayyad, Ingram, Menzies, Ammar & ASE  & 2013 & MOEA & ARE & Stat.Tests & 7,FM$^{1,4}$ \\ \hline

\citeS{DBLP:conf/icse/KarimpourR13} &  Karimpour, Ruhe & ICSE  & 2013 & MOEA & ME & Undefined & 1,FM$^{1}$\\ \hline

\citeS{DBLP:conf/icse/SayyadIMA13} &  Sayyad, Ingram, Menzies, Ammar & ICSE  & 2013 & MOEA & ARE & Stat.Tests & 1,FM$^{1}$\\ \hline

\citeS{DBLP:conf/icse/SanchezMR13} &  Sanchez,  Moisan, Rigault & ICSE  & 2013 & DS & ARE, AR & None & 100,FM$^{2}$ \\ \hline

\citeS{PIC/ShiGW10} &   Shi, Guo, Wang & PIC  & 2010 & GRE & ARE & Basic & 200,FM$^{2}$\\ \hline

\citeS{DBLP:conf/hase/YuDLKK14} &   Yu,Duan, Lei, Kacker, Kuhn & HASE  & 2014 & CH & DT & Basic & 12,FM$^{1}$ \\ \hline

\citeS{DBLP:conf/re/Yi0ZJM12} &    Yi, Zhang, Zhao, Jin, Mei & RE  & 2012 & GA & ME & Basic & 2,FM$^{1}$ \\ \hline

\citeS{DBLP:conf/caise/EnsanBG12} &  Ensan, Bagheri, Gasevic & CAiSE  & 2012 & GA & DT & Basic & 8,FM$^{1}$\\ \hline
 
\citeS{book/TopProductivity/Zhang11} &    Zhang, Haiyan, Mei & BC  & 2011 & DS & DRE & Basic & 32,FM$^{2}$\\ \hline

\citeS{DBLP:conf/icst/HenardPPKT13} &    Henard, Papadakis, Perrouin, Klein,  Le Traon & ICST  & 2013 & LS & DT  & Stat.Tests & 12,FM$^{1,4}$ \\ \hline

\citeS{DBLP:conf/vamos/SeguraGBPC12} &    Segura, Galindo, Benavides, Parejo, Ruiz-Cort{\'e}s & VaMoS  & 2012 & GA & TOOL & None & None \\ \hline

\citeS{DBLP:journals/corr/abs-1211-5451} &    Henard, Papadakis, Perrouin, Klein,  Le Traon & CoRR  & 2012 & GA & DT & Basic & 124,FM$^{1,5}$\\ \hline

\citeS{DBLP:journals/corr/HaslingerLE13} &  Haslinger, Lopez-Herrejon, Egyed  & CoRR   & 2013 & GRE & DT & Stat.Tests & 133,FM$^{1}$ \\ \hline


\end{tabular}
\\ Study type: \textbf{FM} feature model, \textbf{CD} class diagram, \textbf{AH} ad hoc model 
\\ Provenance Superscripts: \textbf{1} SPLOT, \textbf{2} Random, \textbf{3} Open source project, \textbf{4} Academic, \textbf{5} Industrial
\label{tab:primary-sources}
\end{table*}

\subsection{Results}
\label{subsec:results}

\begin{figure}[t]
 \caption{Publications per year since 2001}
 \label{fig:publication-year}
  \includegraphics[scale=0.5]{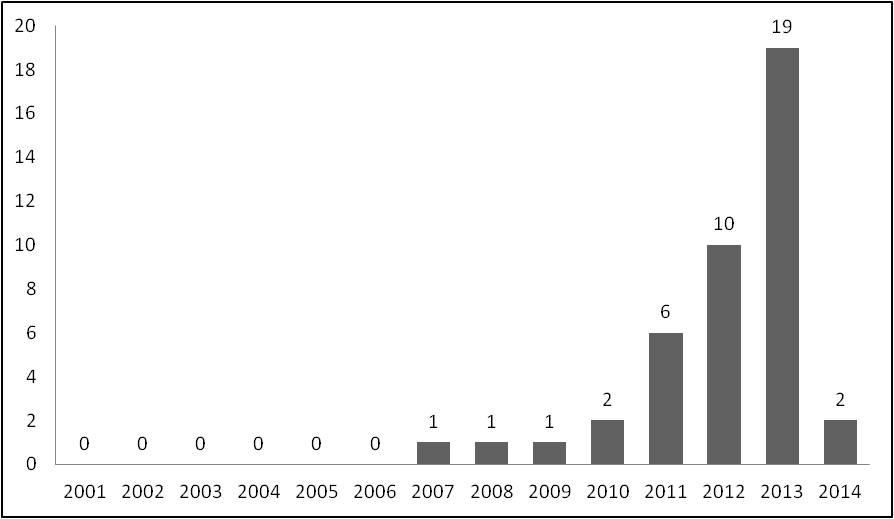}
\end{figure}

The first interesting result of our mapping study is the growth in number  of publications as shown in Figure~\ref{fig:publication-year}. From 2007 to 2009 the growth was steady. Then a sharp increase followed from 2010 to 2013 where each year almost doubled the number of publications  of the previous year. There are early signs that an increasing trend will also continue in 2014.

Regarding the publication fora, we can derive from Table~\ref{tab:primary-sources} that the most popular venue was the ICSE conference with 7 publications, considering short and workshop papers. A close second was the SPLC conference with 6 publications. On third place there was a tie with 3 publications between VaMoS workshop and the Symposium on Search-Based Software Engineering (SSBSE). On fourth place it was Computing Research Repository (CoRR) with 2 papers. The rest of the publications were distributed across 21 distinct venues from journals and conferences to a book chapter (BC) and a technical report (TR).


\subsubsection{Results RQ1 \--- SPL Life Cycle Stages}
\label{subsubsec:rq1}

\begin{figure}[b]
 \caption{SPL Life Cycle Use}
 \label{fig:life-cycle}
  \includegraphics[scale=0.6]{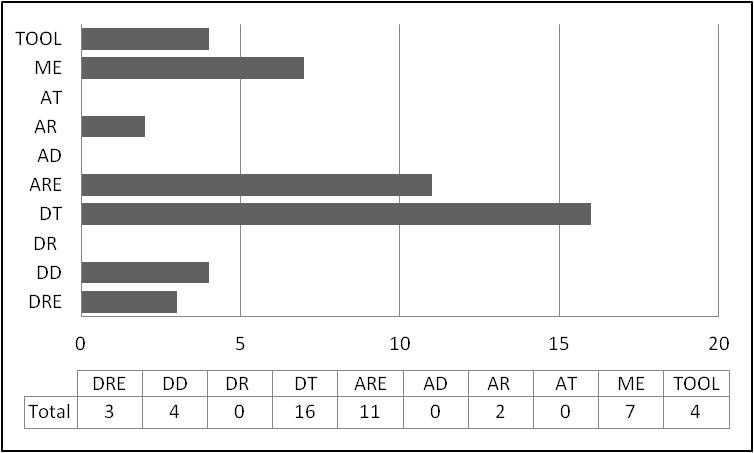}
\end{figure}

Figure~\ref{fig:life-cycle} shows the number of publications classified as described in Section~\ref{subsec:process}. 
The most frequent stage where SBSE techniques are used is DT, testing at Domain Engineering level, with 16 publications. The majority of these publications focus on computing test suites  that cover certain types (e.g. 2-wise or 3-wise) of features combinations that are derived from feature models.
The second place was ARE, requirements engineering in the Application Engineering phase, with 11 publications. Many of these publications dealt with optimizing product configuration or derivation with different characteristics and attributes. Category ME, maintenance and evolution applications, came third place with 7 publications. Among the applications were reverse-engineering and fixing inconsistencies (e.g. of feature models). In fourth place with 4 publications there were TOOL and DD. For the former the tools supported analysis and generation of feature models and visualization, whereas for the latter the focus was on software architecture. 
In fifth place with 3 publications was DRE where the main focus was on the correct definition of variability models. Lastly with 2 publications was AR where the main concern was on runtime adaptation.
 
\newpage
\subsubsection{Results RQ2 \--- SBSE techniques used}
\label{subsubsec:rq2}

Table~\ref{tab:sbse-techniques-used} summarizes what SBSE techniques were used and in how many publications. Not surprisingly genetic algorithms came first with 14 publications. We believe their popularity might be because they are among the most basic evolutionary algorithms. On second place there was the multi-objective evolutionary algorithms with 8 publications. 
We should also point out that we draw a distinction with other forms of multi-objective optimization besides evolutionary, in our study we found two ad hoc exact approaches (AEMOO).
While performing the mapping study, we noticed in some publications a common misunderstanding in distinguishing between genetic algorithms with one fitness function (i.e. single-objective) and multi-objective algorithms. We address this finding in further detail in Section~\ref{sec:guidelines}. 

The third place was greedy algorithms with 7 publications. The fourth place with 4 publications was deterministic searches (e.g. based on breadth first search). 
The fifth place was simulated annealing with 3 publications, followed by local search with 2. The rest of the publications employed an array of different techniques.

\begin{table}
\center
\caption{SBSE techniques used and Frequency}
\begin{tabular}{|c|p{4cm}|c|}
\hline
 \textbf{Acronym} & \textbf{Description} & \textbf{No.} \\ \hline \hline
 MOEA &  Multi-Objective Evolutionary Algorithm  & 8 \\ \hline
 GA & Genetic Algorithm  & 14 \\ \hline
 LS & Local Search & 2 \\ \hline
 IP  &  Integer Programming & 1 \\ \hline 
 GRE & Greedy algorithm & 7 \\ \hline 
 SA    & Simulated Annealing & 3 \\ \hline 
 PSO  & Particle Swarm Optimization  & 1 \\ \hline
 AEMOO & Ad hoc Exact Multi-Objective Optimization &  2 \\ \hline
 ADHS & Ad hoc Heuristic Search & 1 \\ \hline
 DS & Other Deterministic Search  & 4  \\ \hline 
 CH & Constraint Handling & 1  \\ \hline 
 \end{tabular}
\label{tab:sbse-techniques-used}
\end{table}

\subsubsection{Results RQ3. Type of Comparative Analysis}
\label{subsubsec:rq3}
We divided the publications in four categories depending on the type of analysis performed: \textit{i)} \textit{Undefined} whenever we were not able to clearly discern what type of analysis was performed, \textit{ii)} \textit{None} when there was clearly no analysis presented, \textit{iii)} \textit{Basic} when some basic statistical tools were used (i.e. medians, average, standard deviation), and \textit{iv)} \textit{Stat.Tests} when any of the standard statistical analysis test (as sketched in Section~\ref{sec:guidelines}) was used.

The category of \textit{Basic} was the most frequent with 18 publications, followed by \textit{None} and \textit{Statistical Tests} with 9 publications each, and \textit{Undefined} with 6. As mentioned before, because of the randomness involved in most of the SBSE techniques, using the adequate statistical analysis is of utmost importance for the results obtained to be reliable and meaningful. 
Our findings help to raise awareness of the need to employ adequate statistical analysis so that it could be addressed by the SPL community in  large. To contribute to this effort, we outline in  Section~\ref{sec:guidelines} a standard experimental methodology followed when using SBSE techniques.

\subsubsection{Results RQ4. Evaluation Case Studies}
\label{subsubsec:rq4}

We categorize the artefacts employed in the publications in four types: \textit{i)} \textit{None} when there was no clear artefact employed, \textit{ii)} Class Diagrams (\textit{CD}), \textit{iii)} Feature Models (\textit{FM}), and \textit{iv)} Ad hoc (\textit{AH}) when special format artefacts were used in particular cases. 

For the provenance of the artefacts we use five categories: \textit{i)} from the SPLOT repository, \textit{ii)} when artefacts are generated randomly, \textit{iii)} when the artefacts come from open source projects, \textit{iv)} when the artefacts come from academia, and \textit{v)} when the artefacts belong to actual industrial cases.

The trends are clear in both aspects. The great majority of artefacts used are feature models (28 publications), with ad hoc artefacts (that range from specialized constraint formats to cost models) in second place with 8 publications. In terms of provenance, SPLOT was the main source with 19 publications, followed by academic case studies with 9 publications. In third place was random generation with 6 publications, followed closely by open source projects (5 publications), and industrial cases (4 publications).   

An interesting finding was the low number of feature models employed, in some cases as low as 1 or 2 feature models, which raises the issue of how generalizable the results of those studies are.
Another finding is that the majority of publications in the area of domain testing (DT) employed SPLOT feature models. This suggests the possibility of extracting a common benchmark of feature models for assessing the different testing techniques.
 

\subsection{Threats to validity}
\label{subsec:threats}
We faced similar validity threats to any other systematic mapping study. The selection of the search queries was carefully chosen to include common terms in both SPLs and SBSE. For the latter we used an extended the search terms already employed in a survey of SBSE techniques. 
In our search we employed two SBSE paper repositories and 5 standard bibliography search engines.
For our classification terms, we employed as a basis a framework that is well-known in the SPL community. For the data extraction we performed three independent classifications by three groups with different background, which were subsequently validated and consolidated.
Certainly, any of these aspects could be refined and improved to obtained a better snapshot of the application of SBSE to SPLs. We definitely intend to do so with the feedback of both communities.
In addition we will expand our search by looking into the references of the papers already collected.

\section{SBSE Guidelines}
\label{sec:guidelines}

In this section we summarize the common pitfalls and misunderstandings in the use of SBSE techniques for SPL that were revealed by our mapping study. We put forward a set of basic guidelines that researchers interested in applying SBSE techniques can follow to avoid them.

\subsection{Experimental Methodology}

When one wants to compare the performance of a set of algorithms in a task there are roughly two options: provide theorems stating the relative performance of the algorithms or perform an empirical study. Theorems are desirable since they are more general and, provided that their hypotheses hold, the claim will be necessarily true. However, proving such theorems is very difficult (if not impossible) since they require mathematical tools that are not developed or far from trivial. On the other hand, empirical studies can be more easily done, since they only require to run the algorithms over the case studies.

An empirical study is always incomplete and the conclusions we obtain from it can always be biased by, at least,  the selection of the case studies, the parameters of the algorithms, the experiment design and the analysis of the data. It should be clear that the conclusions obtained could be wrong. However, it is possible to reduce the probability of claiming wrong conclusions. In particular, we will focus in this section on how  to apply an appropriate experimental methodology and statistical analysis of the results to increase the confidence on the conclusions.

One of the first things we should think about when designing the empirical study is how many and what case studies we will use. If the number of case studies is too low (one or two) the conclusions cannot be generalized, there is not enough evidence to claim that the results will be the same if new case studies are added to the empirical study. A serious empirical study should contain a large number of case studies. The exact number will depend on the availability of case studies or the budget limit to do the experiments. We should also take into account the No Free Lunch Theorem~\cite{Wolpert1997}, which claims that all the algorithms perform the same when all the problems are considered. An empirical study is focused on one single problem (not all of them), but even in this case the Focused No Free Lunch Theorems~\cite{Whitley2008} suggest that we could find that all the algorithms perform the same as the number of case studies increases. 

If stochastic algorithms are used in the experimental study then the performance measure of the algorithm for a case study is not a single number, but a probability distribution. The result of a single run of the algorithm contains little information about this distribution and several independent runs should be done. Arcuri and Briand~\cite{Arcuri2014} suggest 1,000 independent runs of the algorithm. There is, however, no reason to use exactly this number, it usually depends on the time limit for doing the experiments. The larger the number of independent runs, the higher the confidence we have on the results. In the case of deterministic algorithms, where the result of the performance measure does not change in different runs, one single execution of the algorithm is enough. If the performance measure is time, even a deterministic algorithm would require several runs, since the wall clock time depends on the load of the system in which the algorithm runs.

Once all the results of the different independent runs of the algorithms over the case studies are collected, a statistical analysis of the data is necessary. Common questions we want to answer in this phase are: Do the algorithms perform all the same?, Which is the best algorithm for this task? Data contain several values for the performance of the algorithms on the case studies (one for each independent run). Aggregating these values into a single one, e.g., the average, to compare the algorithms is not enough, since the aggregated value (whatever it is) is a random variable itself. 
For example, if the average execution time of algorithm A for case study X is lower than the average execution time of algorithm B we cannot claim that A is faster than B, since it could be a matter of chance. 
It is necessary to apply statistical tests to check if the observed differences are really significant or not. There are many statistical tests that can be applied depending on the kind of data and the assumptions that can be made on them. The work of Sheskin~\cite{Sheskin07} is a good reference to find the appropriate test. In short, any statistical test formulates a base hypothesis, called the \emph{null hypothesis} $H_0$, and it computes the probability of having the observed data provided that $H_0$ is true. This computed value is the so-called $p$-value. If the $p$-value is low enough we can safely reject the null hypothesis, meaning that probably it is not true. In the example of the execution time the null hypothesis could be $H_0$: ``the average of the execution time of algorithms A and B over case study X is the same''. Let's say that an appropriate statistical test (a $t$-test, for example) provides a $p$-value of $0.01$. Then we can claim with significance level $\alpha=1\%$ that the mean execution time of algorithm A over all the possible runs (potentially infinite) is lower than the mean execution time of algorithm B. The probability of failing in our conclusion, that is, making a type I error, is $0.01$.

Statistical tests can be classified into parametric and non-parametric. The former assume that the data are samples of a given distribution (e.g., normal, binomial, etc.). One salient example of parametric tests is the Student $t$-test. This test assumes a normal distribution of the random variables and the same standard deviation in both populations. In general, the probability distribution of the data we collect is unknown, so it is difficult to justify the application of parametric tests. In other cases, the assumptions of the parametric tests are clearly violated. Taking again the example of the run time of the algorithms, the time does not follow a normal distribution because it cannot be negative, while a normal distribution would require a nonzero probability of having negative values. If the assumptions are violated or it is unknown if they are, a non-parametric test should be used. These tests do not assume a distribution of the data, so they can always be  applied. A popular non-parametric test to compare two samples is the Mann-Whitney U-test, which checks of the median if the ranks of the samples are equal or not.

Care should be taken if many statistical tests are done during the statistical analysis. For example, it is common to compare many algorithms by applying two-samples tests for all the possible pairs of algorithms. In this case, the global $p$-value obtained of the experiment is higher than the $p$-values for each particular comparison. The probability of type I errors are accumulated. Thus, using a significance level of $\alpha$ in each pairwise comparison is not the same as having a significance level of $\alpha$ in the global experiment, the latter is higher, and this could affect the confidence on the conclusions. In order to bound the significance  level of the global experiment by a known value, it is required to reduce the significance level of the pairwise comparisons. There are many proposals for this. One well-known approach is the Bonferroni correction~\cite{Nakagawa2004}.

\subsection{Multi-objective Optimization}

In our survey we found two main kinds of multi-objective techniques: the ones that use a mono-objective technique with a weighted sum of the objectives as fitness functions (e.g.~\citeS{DBLP:conf/gecco/WangAG13, DBLP:conf/splc/HenardPPKT13}) and the ones that use multi-objective algorithm to find the entire Pareto front. The former technique has several drawbacks. If the weights of the aggregative function are not systematically varied during the optimization, a single trade-off solution is obtained. In addition, even if the weights are changed during the search, it is not possible to obtain all the points in the Pareto front if it is concave downwoards~\cite{Das1997}. The use of weighted sum of objectives could be useful if the preferences of all the objectives are clear. When this is not the case, multi-objective algorithms should be used, since they are able to obtain the Pareto front~\cite{CoelloCoello2007}. These algorithms provide an approximated Pareto front, which is a set of non-dominated solutions. One important issue is how to compare these fronts, since they are not just a single number. 
This is the role of quality indicators as explained next.

\subsubsection{Quality Indicators}

Three different issues are normally considered for assessing the quality of the results computed by a multi-objective optimization algorithm~\cite{Zitzler2000}:
%
\textit{i)} To minimize the distance of the computed solution set by the proposed algorithm to the optimal Pareto front (convergence towards the optimal Pareto front),
\textit{ii)} To maximize the spread of solutions found, so that we can have a distribution as smooth and uniform as possible (diversity),
and
\textit{iii)} To maximize the number of elements of the Pareto optimal set found.

A number of quality indicators have been proposed in the literature trying to capture the three issues indicated above, but for the moment, there is not a single metric which captures all of them. Consequently, researchers should use more than one to measure different aspects of the solutions generated by the multi-objective techniques. Among them, we can distinguish between \textit{Pareto compliant} and \textit{non Pareto compliant} indicators~\cite{KTZ06}. Given two Pareto fronts, A and B, if A dominates B, the value of a Pareto compliant quality indicator is higher for A than for B; meanwhile, this condition is not fulfilled by the non--compliant indicators. Thus, the use of Pareto compliant indicators should be preferable. To apply these quality indicators, it is usually necessary to know the optimal Pareto front. However, the location of the optimal front is usually unknown. 
Therefore, the front composed of all the non--dominated solutions computed by all analyzed approaches is used to obtain a reference Pareto front. 
Many quality indicators have been proposed in the literature. Next we highlight the advantages and disadvantages of some of the most common ones:

\textbf{Number of Pareto optimal solutions}. This non--compliant indicator is very simple, it computes the number of solutions that are included in the optimal Pareto front. Its main advantage lies in the fact that it is very easy to compute. In contrast, the disadvantages are the lack of information about the diversity of solutions and the requirement of knowing the optimal Pareto front.

\textbf{Hypervolume (HV)~\cite{ZT99}.} This Pareto--compliant indicator calculates the volume (in the objective space) covered by members of a non-dominated set of solutions. For each solution of the set, a hypercube is constructed with a reference point and the solution as the diagonal corners of the hypercube. 
The main advantages of the hypervolume are that it considers the convergence as well as the diversity of the solutions, and it doesn't require the optimal Pareto front. A drawback is that it depends on the reference point selected. Different reference points produce different results. This could be critical to compare the results with existing approaches in the literature. Therefore, it should be always reported what is the reference point used for computing the hypervolume.

\textbf{Spread (SD)~\cite{Deb2001}.} It is a diversity quality non--compliant indicator that measures the distribution of individuals over the non-dominated region. This measure is based on the distance between solutions, so Pareto fronts with a smaller value of Spread are more desirable. 
The main advantage of this measure is that it summarizes the diversity of a Pareto front in one single scalar value. The main disadvantage is that it does not consider the other two quality aspects, i.e.,  the solution set could be very well distributed, but the solutions could be far from the optimal Pareto front. Other quality indicators should be used to complement the Spread.

\textbf{Generational Distance (GD)~\cite{VanVeldhuizen:1999:MEA:929368}.} The generational distance is a non--compliant indicator. It measures how far the elements in the approximated Pareto front are from those in the optimal Pareto front. It considers the distance of the approximated Pareto front obtained to the reference front.
Pareto fronts with a smaller value of GD are more desirable.
The advantages of the generational distance are the ease of understanding and calculation, and the possibility to use different kinds of distance functions. In contrast, it does not take into account the diversity of the solutions found, i.e., a front with only one solution in the optimal Pareto front will obtain an ideal value of generational distance.

\textbf{Epsilon (Multiplicative)~\cite{Zitzler2003}.} This Pareto--compliant indicator measures, in one single scalar value, how badly approximated the worst approximated solution of the Pareto front is. The approximation quality of solutions is the ratio between the optimal value and the best value found. 
The main advantage of this quality indicator is that it allows us to compare the quality of solutions between different functions, different population sizes, and even different dimensions. In addition, it measures convergence of the algorithm, but it does not depend on a chosen reference point like the hypervolume. In contrast, its main disadvantage is that it only considers part of the front, namely the worst solution.


The previous indicators have the advantage of summarizing an entire front into one single scalar value that allows the performance of different algorithms to be compared. However, from the point of view of a decision maker, knowing about a single number is not enough, because it gives no information about the shape of the front.

In the related literature, the trade-off between the different objectives is usually presented by showing one of the approximated Pareto fronts obtained in one single run of a given algorithm. 
However, if the optimization algorithm used is stochastic there is no warranty that the same result is obtained after a new run of the algorithm. 
We need a way of representing the results of a multi-objective algorithm that allows us to observe the expected performance and its variability, in the same way as the average and the standard deviation are used in the single-objective case. For this reason, the concept of \textit{Empirical Attainment Function (EAF)}~\cite{knowles05summary} is used. In short, the EAF is a function $\alpha$ from the objective space $\mathbb{R}^n$ to the interval $[0,1]$ that estimates for each vector in the objective space the probability of being dominated by the approximated Pareto front of one single run of the multi-objective algorithm. Given the $r$ approximated Pareto fronts obtained in the different runs, the EAF is defined as:
\begin{equation}
\alpha(z) = \frac{1}{r} \sum_{i=1}^{r}I(A^i \preceq \{z\})
\end{equation}
where $A^i$ is the $i$-th approximated Pareto front obtained with the multi-objective algorithm and $I$ is an indicator function that takes value 1 when the predicate inside it is true, and 0 otherwise. The predicate $A^i \preceq \{z\}$ means $A^i$ dominates solution $z$. Thanks to the attainment function, it is possible to define the concept of $k$\%-attainment surface~\cite{knowles05summary}. The attainment function $\alpha$ is a scalar field in $\mathbb{R}^n$ and the $k$\%-attainment surface is the level curve with value $k/100$ for $\alpha$.

Informally, the 50\%-attainment surface is analogous to the median in the single-objective case. In a similar way, the 25\%- and 75\%-attainment surfaces can be used as the first and third ``quartile fronts'' and the region between them could be considered a kind of ``interquartile region''.

The attainment surfaces provides engineers with a tool for evaluating the variability of an algorithm for the problem at hand.
The variability in the results of one multi-objective algorithm is not reduced to a scalar (as in the single-objective case). This variability depends on the region observed in the objective space. We can find a rich range of possibilities when considering variability in the multi-objective domain. Using attainment surfaces the engineer can analyze and explore this range of possibilities. From a practical point of view, in the SPL domain, this tool helps the engineer to decide on the more suitable multi-objective algorithm for her/his requirements.

\section{Related Work}
\label{sec:related-work}

In this section we briefly summarize the salient surveys and studies carried out in either SPLs or in SBSE. 
The survey by Harman et al. presents a general overview of SBSE techniques and the areas where it has been employed~\cite{DBLP:journals/csur/HarmanMZ12}. Freitas et al. performed a bibliometric analysis of SBSE~\cite{DBLP:conf/ssbse/FreitasS11}. Their goal was to identify trends in the number of publications, the publication fora, the authorship and collaborations amongst members of the SBSE community.
In contrast with our work, they have a different focus, namely general software engineering. 
Ali et al. performed a systematic review of empirical investigation of search-based test case generation techniques 
~\cite{DBLP:journals/tse/AliBHP10}. Similar to ours their review assessed how the focused techniques were empirically evaluated, but in contrast their work focused exclusively on testing and for non-SPL software systems.

In the area of SPL there are two recent systematic mapping studies in SPL testing~\cite{DBLP:journals/infsof/EngstromR11,DBLP:journals/infsof/NetoMMAM11}, both of which attest that the application of SBSE techniques for SPL testing is an area ripe for research that needs to be further explored.
Laguna et al. performed a systematic mapping study on SPL evolution~\cite{DBLP:journals/scp/LagunaC13}, where they made an assessment of the maturity level of techniques to migrate individual systems or groups of software variants into SPLs. Rabiser et al. performed a systematic review of requirements for supporting product configuration~\cite{DBLP:journals/infsof/RabiserGD10}, whereas Holl et al. carried out a systematic review of the capabilities to support multi product lines~\cite{DBLP:journals/infsof/HollGR12}. Chen et al. performed a systematic review of variability management~\cite{DBLP:journals/infsof/ChenB11}. In contrast with all these SPLs studies our work has a novel and distinct focus.

\section{Conclusions and Future Work}
\label{sec:conclusions}

In this paper we present the results of the first systematic mapping study on the application of SBSE techniques to SPL problems. Our study corroborates the increasing interest in applying this type of techniques as shown by the  number of recent publications. 
The most common application is for testing at the Domain Engineering level, for example in computing test suites in combinatorial interaction testing. 
The most common technique used is genetic algorithms with an increasing interest in multi-objective optimization problems. 
We identify a need to improve empirical evaluations with a more adequate statistical analysis, and some common pitfalls when dealing with multi-objective optimization algorithms. 
To address these two issues we provide a short guideline section to serve as a reference basis for researchers and practitioners interested in exploiting SBSE techniques. 

Our work also revealed research areas and possible opportunities. For example, we inadvertently found that there is a plethora of work on product line scoping and design in the area of manufacturing and marketing that relies on SBSE techniques. This begs the question if any of the research done in those areas can be applicable to SPLs. 
We found no applications in domain realisation, application design, and application testing. For the first two there is work in SBSE on the generation of software artefacts including architectural models that could be leveraged, the challenge is how to effectively express and cope with variability issues.


%
\section{Acknowledgements}

This research is partially funded by the Austrian Science Fund (FWF) projects P25289-N15, P25513-N15, and Lise Meitner Fellowship M1421-N15, the Spanish Ministry of Economy and Competitiveness and FEDER under contract TIN2011-28194 and fellowship BES-2012-055967. It is also partially founded by
project 8.06/5.47.4142 in collaboration with the VSB-Technical University of Ostrava and Universidad de M\'alaga, Andaluc\'{\i}a Tech.



\bibliographystyle{abbrv}
\bibliography{splc14-biblio}

\appendix
\bibliographystyleS{abbrv}
\bibliographyS{splc14-primary-resources}

\end{document}